# Interlayer Water Regulates the Bio-nano Interface of a β-sheet Protein stacking on Graphene

*Wenping Lv[a], Guiju Xu[a], Hongyan Zhang[a], Xin Li[a], Shengju Liu[a], Huan Niu[a], Dongsheng Xu[b], Ren'an Wu\*,[a]*

[a] CAS Key Laboratory of Separation Science for Analytical Chemistry, National Chromatographic R & A Center, Dalian Institute of Chemical Physics, Chinese Academy of Sciences (CAS), Dalian 116023, China

[b] Institute of Metal Research, Chinese Academy of Sciences, Shenyang 110016, China

\* Corresponding Author

**Contact information:**

Renan Wu

Email: wurenan@dicp.ac.cn

Wenping Lv

Email: lv_wenping@live.cn; wenping@dicp.ac.cn

**KEYWORDS**: water, protein, graphene, interface, interaction

# ABSTRACT

Using molecular dynamics simulations, we investigated an integrated bio-nano interface consisting of a β-sheet protein stacked onto graphene. We found that the stacking assembly of the model protein on graphene could be controlled by water molecules. The interlayer water filled within interstices of the bio-nano interface could suppress the molecular vibration of surface groups on protein, and could impair the CH···π interaction driving the attraction of the protein and graphene. The intermolecular coupling of interlayer water would be relaxed by the relative motion of protein upon graphene due to the interaction between water and protein surface. This effect reduced the hindrance of the interlayer water against the assembly of protein on graphene, resulting an appropriate adsorption status of protein on graphene with a deep free energy trap. Thereby, the confinement and the relative sliding between protein and graphene, the coupling of protein and water, and the interaction between graphene and water all have involved in the modulation of behaviors of water molecules within the bio-nano interface, governing the hindrance of interlayer water against the protein assembly on hydrophobic graphene. These results provide a deep insight into the fundamental mechanism of protein adsorption onto graphene surface in water.

Bio-nanotechnology has advanced to the point at which it can now focus on understanding and controlling bio-nano interface interactions.[1-4] The basic biomolecule and the star nanomaterial, protein and graphene, play significant roles in the revolution of science and industry. The bio-nano interface between a protein and graphene has attracted broad interest for the development of biocompatible hybrid materials, nanomedicines, biomimetic sensors, and protein separation and purification strategies.[5-10] Because water is the universal solvent in biological systems, it is involved in the formation of bio-nano interfaces. Understanding the interactions between a protein and graphene in a water environment is fundamental to our ability to regulate protein and graphene interactions.

Investigations on protein-nanomaterial interfaces in aqueous solutions have progressed from the colloidal scale to the molecular level.[11-13] Water has been found to actively participate in the assembly of proteins on a solid surface by mediating interactions between binding partners and protein conformations.[12,14-17] Based on molecular dynamics (MD) simulations and statistical analyses, it has recently been proposed that the orientations and ordering of water layers adjacent to an uncharged solid surface dictate the mechanism of adsorption for a protein onto an uncharged solid.[12] Near the surface of graphene, hydrophobic hydration results in hydrogen bond (HB) fluctuations, density oscillations and dipole biasing of water molecules.[18-22] Non-covalent interactions (NCIs), such as van der Waals (vdW) and π-π stacking between proteins and graphene,[23-26] as well as heterogeneous and dynamic hydration environments impact the assembly of proteins on graphene;[12,14,27,28] *e.g.* the *α*-helix structure of a peptide on a dry graphene surface can transform into a distorted helical structure in the presence of water.[14] In addition, the flexibility of a graphene monolayer has been shown to be important for the adsorption of small peptides and for the depletion of water interlayer.[27,29] Understanding the interactions between protein, water and graphene is a key to controlling their behaviors within bio-nano interfaces.[14,24,27,28] Unfortunately, the interactions within a water-regulated protein-graphene (P-G) interface, particularly the molecular behavior of water within the assembled interface between a bulk protein and graphene, have not yet been well characterized.

In this contribution, we demonstrate the molecular details of the protein, water and graphene interactions, as well as the role of water in the P-G interface, by analyzing hundreds of all-atom MD simulations. The model protein is a planar β-sheet antifreeze protein (AFP) containing regular outward-projecting parallel arrays of threonine (Thr) residues;[30] these regular arrays allow for uniform protein affinities across the bio-nano interface. The secondary structure of this model protein allows it to bind onto several abiotic surfaces, including graphene, without damage.[14,15,26,31] Furthermore, the planar shape of this model protein allows for the formation of an integrated P-G interface to investigate the role of interlayer water within the bio-nano interactions. We found that not only the water molecules could hinder the interactions of the model protein and graphene surface but also the transversal motion of the protein would reduce this hindrance simultaneously. We explored the mechanisms of the formation of bio-nano interface with the regulation of water in aspects of molecular bonds vibrations, NCIs, free energy profiles, and hydrogen bonds dynamics, to provide a comprehensive understanding of the interplay among protein, water and graphene in this interface.

# RESULTS

*Interlayer water modulates the P-G interaction of β-sheet protein stacking on a graphene surface.*

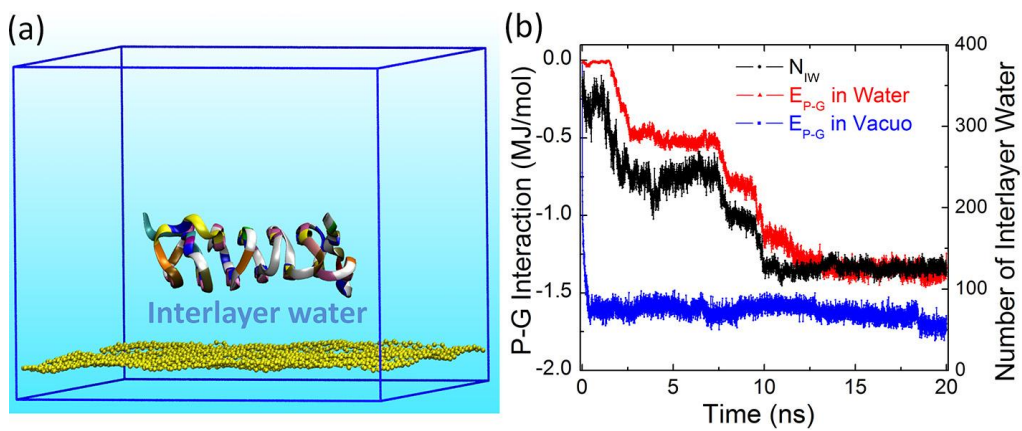

**Figure 1** (a) A initial structure of the simulation system. Protein and graphene were drawn by the methods of NewCartoon and CPK, respectively. (b) P-G interaction over time in vacuo, and in water with respect to number of interlayer water.

AFP self-assembly on a graphene surface has been performed to create a bio-nano interface consisting of a β-sheet protein stacked on graphene (*MD Indexes 1-2*, Supplementary **Table S1**). The initial orientation of the protein is adjusted to be parallel with the graphene surface (**Figure 1a**) with a small separation of approximately 1.5 nm to ensure that a stacking conformation exists between the model protein and graphene upon aggregation. As expected, the water molecules in the interval between the protein and graphene, called as the interlayer water, were able to effectively retard the stacking of two planar units. In above simulations, the stacking of AFP on graphene in water would require a prolonged period of time (*ca.* 10 ns) as compared to the rapid aggregation (*ca.* 0.5 ns) of AFP on graphene in vacuo (**Figure 1b**). In water, the aggregation underwent stages of biased diffusing (2 ns) of AFP, touching between AFP edge and graphene (5 ns), draining of the interlayer water (3 ns), and stacking of AFP on graphene with the residual interlayer water (**Supplementary Movie S1**). Interestingly, the amount of interlayer water ($N_{IW}$) coincided with the undulations of P-G interaction ($E_{P-G}$) (**Figure 1b**), with a linear relationship between $N_{IW}$ and $E_{P-G}$ (**Supplementary Figure S1**). Even after the protein was stacked onto graphene (the last 10 ns of the simulation), over 100 water molecules were retained in the interlayer. Correspondingly, as shown in Figure 1b, the P-G interaction was greater than -1,500 kJ/mol in vacuo ($E_{P-G}$ in vacuo) but decreased to approximately -1,250 kJ/mol in water ($E_{P-G}$ in water) after the 10 ns. Herein, the negative sign of P-G interaction represented only an attraction between protein and graphene. Additional simulations (*MD Indexes 3-4*) demonstrated that the presence of interlayer water was independent on the water models employed (**Supplementary Figure S2**), firming up a persistently existence of these interlayer water molecules in the formation of a stacked P-G interface.

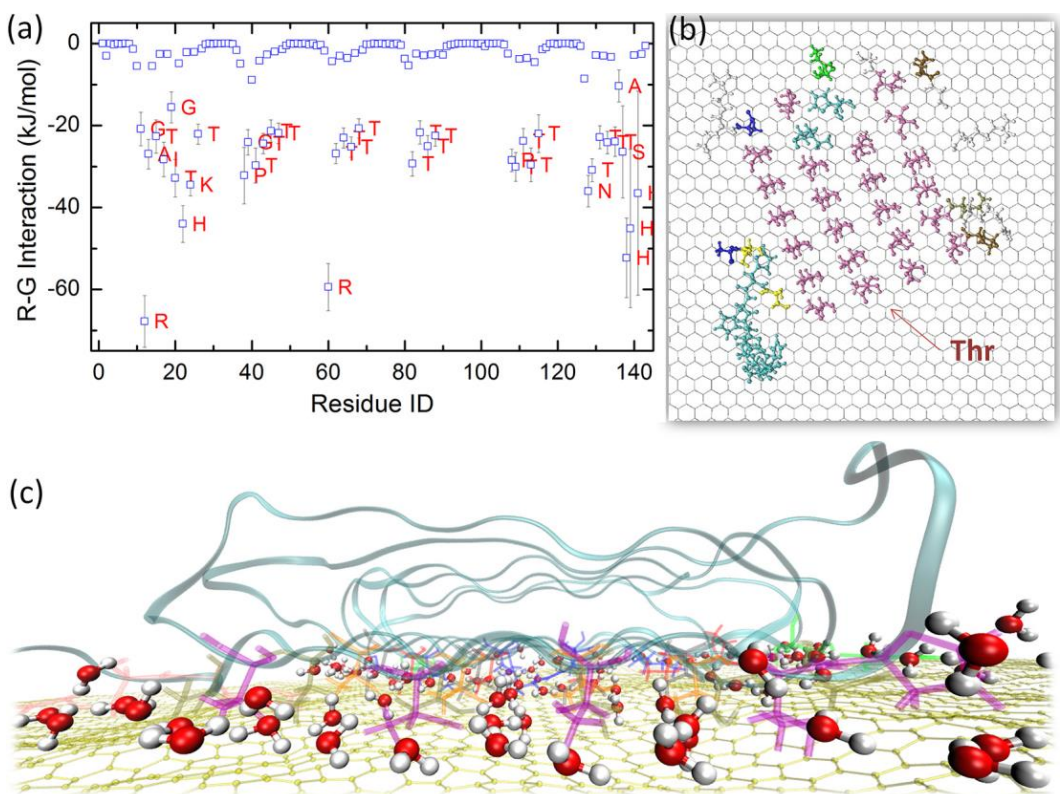

**Figure 2** (a) Average R-G interaction in water during the last 10 ns of simulation. The fluctuations (error bars) of the R-G interaction were also plotted for contributing residues those tagged with their single letter abbreviations. (b) The locations of the contributing residues (tagged in (a)) on graphene surface. The Thr arrays dominate most of the contact region inside the interface. (c) A side view of the P-G interface with the interstices around the contributing surface residues (shown in Licorice model) filled with the interlayer water molecules (shown in CPK model).

To further evaluate the P-G interaction, the average interactions for each protein residue and graphene (R-G interaction, $E_{R-G}$) during the last 10 ns of the simulation were investigated (**Figure 2a**). Results showed that the contributing residues ($|E_{R-G}| > 12.5$ kJ/mol, tagged with their single letter abbreviations) were distributed on the surface of the protein, and the regular Thr arrays were the main components (**Figure 2b**). Interestingly, the water molecules in the interlayer could not prevent contact between surface groups and graphene directly, but rather filled in the interstices around the contributing residues of the adsorbed-AFP (**Figure 2c**). A detailed discussion on the difference of R-G interactions in vacuo and in water (**Supplementary Text**) showed that the interlayer water molecules could impair the R-G interactions of the contributing residues those inside the P-G interface.

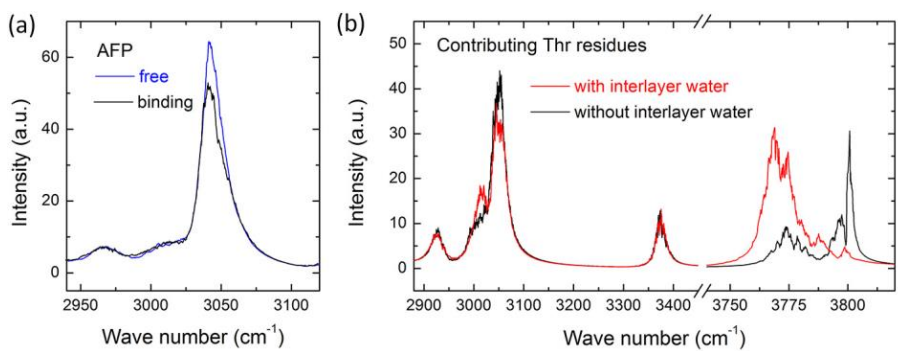

**Figure 3** (a) The VPS corresponding to C-H bond stretching of the methyl groups on Thr residues (free and binding to graphene) in water. (b) The VPS of contributing Thr residues on binding-AFP surface (with or without interlayer water).

The vibration power spectra (VPS) of the AFP and the contributing Thr residues on protein surface were investigated respectively, to probe the influences of water molecules in the interlayer on the dynamics behaviors of the adsorbed-protein. Firstly, we compared the vibrational change of AFP induced by P-G adsorption (**Figure 3a**), showing that the VPS was weakened approximately 10% (peak area, from 3020 to 3080 cm$^{-1}$) after AFP was adsorbed onto graphene in water. According to infrared spectra and theoretical calculations[32,33] this vibration mode could be a result of stretching vibrations within C-H molecular bonds of methyl groups (**Supplementary Figures S2-S3**), suggesting that the interaction between the protein and graphene restrained the molecular bond vibrations of protein surface groups. Then, the VPS of the contributing Thr resides (**Figure 3b**) inside the P-G interface were calculated to confirm the influence of the interlayer water on the behavior of these surface groups. Here, an additional simulation was performed (*MD Index 13, Supplementary Table S1*), in which the interlayer water had been removed manually. Results showed that the filling of water molecules in the interlayer weakened the stretch vibration of C-H molecular bonds (*ca.* 3050 cm$^{-1}$), and induced an enhancement of small peak around 3020 cm$^{-1}$. It means that the inhibited molecular vibration of the protein stacking on graphene (**Figure 3a**) was caused by not only the P-G interaction but also the hindrance of interlayer water. In addition, the vibration modes around 3775 cm$^{-1}$ were reduced significantly, and shifted to 3800 cm$^{-1}$ when the interlayer has no water molecule. Such an obvious frequency shift implied a relatively strong interaction between the interlayer water and Thr residues.

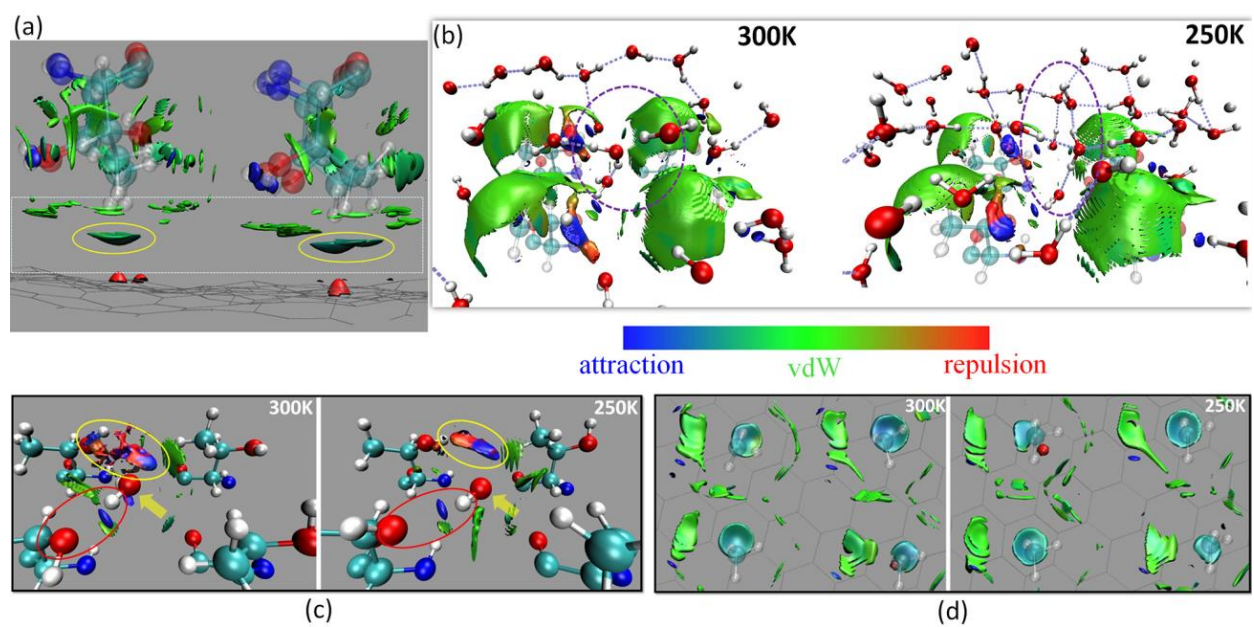

**Figure 4** (a) An isosurface side view of the average RDG of the P-G interface. The white dot-line box highlights the NCIs between the Thr residues and graphene, and the yellow circles marks the CH···π attractions between the methyl groups of Thr and graphene. (b) The average RDG of water-AFP interfaces at 300 and 250 K. The same region monitored in (a) is monitored here. Four Thr residues, water molecules near four Thr residues (< 0.5 nm) and HB networks are displayed in the transparency CPK model, the CPK model, and as dotted lines, respectively. Graphene is not shown for clarity. (c) Attraction coupling (blue isosurface within the yellow circle) between a protein-bound water molecule (marked with a yellow arrow) and surrounding unbound water molecules (not shown) at 300 and 250 K. Red circles highlights the attraction between Thr residues and protein-bound water molecules. (d) The CH···π attraction between Thr residues and graphene at 300 and 250 K. The -CH$_3$ groups on Thr residues are drawn in the transparent CPK model to show the contact points between protein and graphene. The value of above displayed isosurfaces is 0.3. Blue, green and red colors on these isosurfaces represent attraction, vdW and repulsion NCIs, respectively.

To understand NCIs within such a water filled bio-nano interface between protein and graphene visually, the reduced electron density gradients (RDGs) between stacked-AFP and graphene were subsequently studied using electron structures constructed by promolecular approximation.[34-36] Considering thermal fluctuations in the environment,[36] 14,500 structures were sampled to obtain average RDG isosurfaces between protein and graphene after a relaxation time of 11 ns. Because Thr was the most contributing residue in the R-G interaction (**Figure 2a**), four Thr residues at the center of the P-G interface were monitored (**Supplementary Figure S5**). The color of isosurfaces of the obtained average RDGs represents the nature of the NCIs.[34-36] As shown in the white box of **Figure 4a**, several weak interactions exist between these Thr residues and graphene. That is, except for the attractions (colored in blue-green) between Thr methyl groups and graphene (CH···π interactions,[37] highlighted with yellow circles), the NCIs between the protein and graphene are primarily scattered vdW interactions (colored in green), similar to the weak surface/side-chain interactions between polyserine and Au(111).[15,38] However, this P-G bonding is distinct from the incipient oxygen-to-gold dative bonding between a β-sheet and a gold surface[15]; the physical nature of the CH···π interaction and the vdW interactions are similar here because both are attributed to dispersion interactions.[37]

Moreover, the intermolecular coupling of water in the interlayer and its influence on the CH···π interaction between Thr residues and graphene were investigated (left panel of **Figure 4b**). The green isosurfaces show the vdW interactions between the interlayer water molecules and the methyl groups of Thr residues. Meanwhile, some of water molecules embedded in the interstices among the four Thr residues displayed strong attractions (small blue isosurfaces) to the surface groups of protein. We further rendered the average RDG between the embedded water molecule and the Thr residues in the left panel of **Figure 4c**. The isosurfaces highlighted with red circles show that embedded water molecules (marked with yellow arrows) were bound to protein via stable HBs with hydroxyl groups on Thr residues. Such an intrinsic coupling difference between protein and embedded water and protein and interstitial water molecules demonstrates that there were two types of water molecules (protein-bound and unbound) within the P-G interface. This finding implies that the draining of unbound water molecules from the P-G interface could be much easier than that of the protein-bound water molecules.

At 300 K, a motley RDG isosurface was observed between the two types of water molecules (within yellow circle of left panel of **Figure 4c**). By cooling the system to temperature of 250 K from the same stacking conformation of protein on graphene at 300 K, an extended blue region of the RDG isosurface was observed between the protein-bound and unbound water molecules (right panel of **Figure 4c**), indicating a strong attractive interaction between water molecules within the interlayer. As a result, abundant HBs appeared among these interlayer water molecules (highlighted with the dotted circle in right panel of **Figure 4b**). Interestingly, with the enhancement of the HBs between the protein-bound water and the unbound water molecules inside the P-G interface, the CH···π attraction between the methyl groups and graphene were also reduced, as the color of the corresponding isosurfaces (blue-cyan) became hypochromic when temperature was changed from 300 K to 250 K (**Figure 4d**). The transition of the NCIs inside the water-filled P-G interface indicates that the modulate role of the interlayer water in P-G interaction can be attributed to the intermolecular coupling of those water molecules in the interlayer.

*Transversal protein migration reduces the AFP-graphene hindrance caused by the interlayer water.*

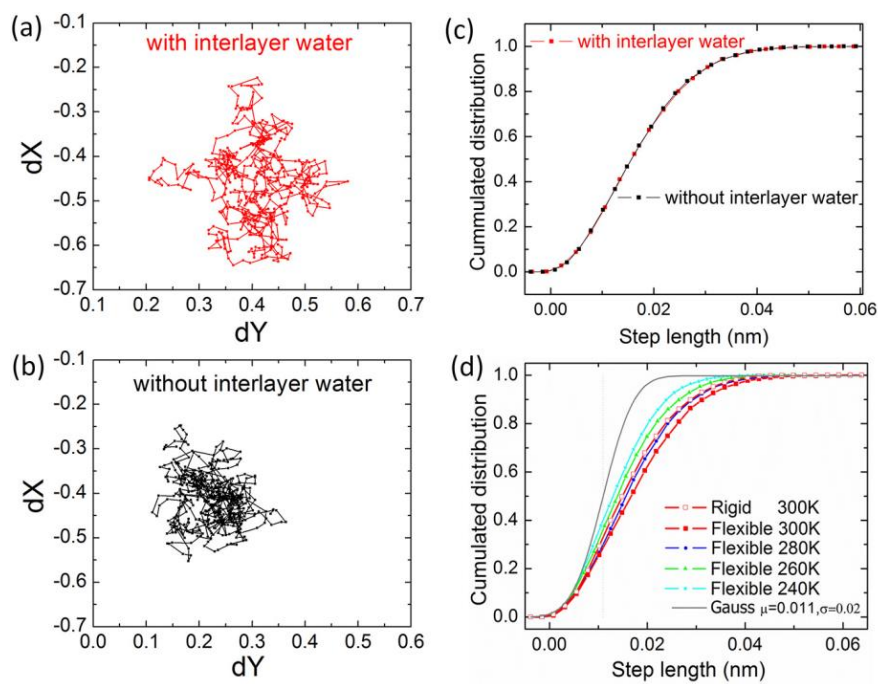

**Figure 5** (a-b) The trajectories of the adsorbed-AFP migrated on graphene with (a) and without (b) the existence of interlayer water (only the data of the first 1 ns in the two 10 ns MD simulations was showed for clarity). (c) A comparison of the cumulated distributions of the migration step length of the adsorbed-AFP with and without the existence of interlayer water. (d) The heavy-tailed distributions of the migration step length for AFP adsorbing onto a graphene surface at different temperatures compared with a Gaussian distribution. "Rigid" means that all atoms of graphene were restrained during simulations; "Flexible" means that the graphene was rippled during simulations.

Other than the influence of the interlayer water on the P-G interaction, the dynamics of protein on graphene were further investigated. We plotted the trajectories of protein (mass center) migration on graphene with and without the presence of interlayer water, to show that the AFP walked in a larger range on graphene with the existence of interlayer water (Figure 5a) as compared to the situation of no interlayer water (Figure 5b). It means that water molecules confined inside the P-G interface could promote the sliding of the graphene-adsorbed AFP. However, two heavy-tailed distributions of the migration step length of protein were observed nearly same whether the interlayer water exists or not (**Figure 5c**), demonstrating a Levy flight [39] characteristic of such an unanchored adsorption of AFP on graphene. To exclude the influence of the ripple of graphene surface, a rigid graphene was modeled by applying a harmonic restraint on all atoms of graphene. The heavy-tailed distribution of the migration step length of protein was observed on either a flexible (rippling) graphene surface or a rigid (flat) graphene surface, but the width of the distribution became narrower for the rigid graphene (**Figure 5d**), suggesting that the rippling surface of graphene could also promote the migration of AFP. A rigid graphene means that the thermal motion of carbon atoms is less violent as compared to the flexible graphene. This effect could influence the thermodynamics of water molecules around graphene, implying that the thermal motion of the interfacial water molecules on graphene determined the Levy flight behavior of protein. By cooling the system, the migration range of the adsorbed-AFP on graphene surface was decreased gradually from 300 K to 280, 260, and 240 K (**Supplementary Figure S6**). The "heavy tail" of the step length distributions was reduced at lower temperatures although the water-promoted sliding of AFP on graphene always showed a Levy flight characteristic as compared to a Gaussian distribution with a central step length (μ) equal to the average value (approximately 0.011 nm) of the most accessible step length of four tested temperatures. The decreased step length distribution of AFP migration at low temperatures confirmed that it was the decreased-thermal motion of water molecules (diffusion constants of water were 3.42, 2.17, 1.29, and $0.43 \times 10^{-5}$ cm$^2$/s for the 300, 280, 260, and 240 K, respectively) dominated the character of protein migration on graphene, meaning that the unanchored adsorption of protein on graphene surface was a entropy effect induced by water.

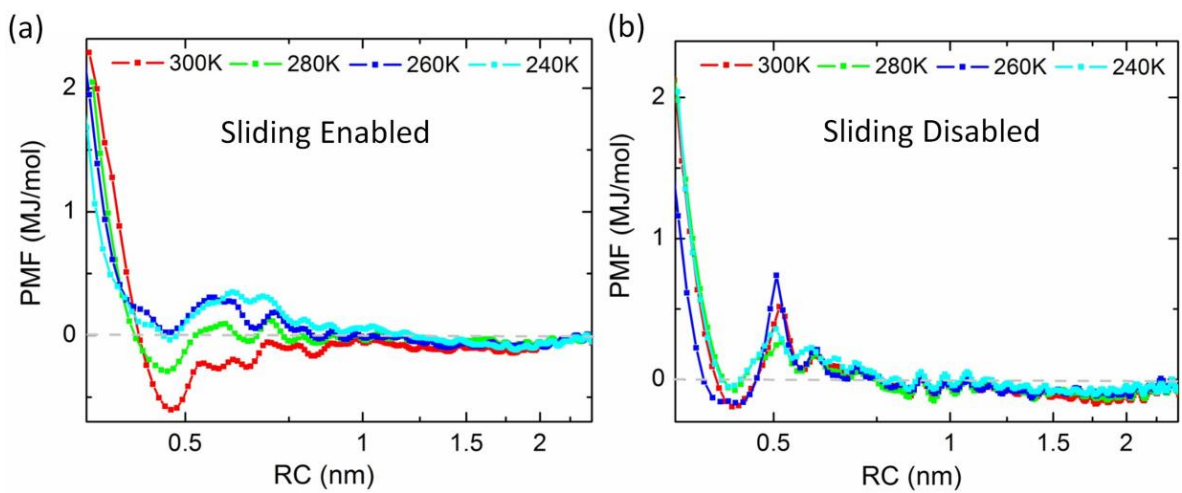

**Figure 6.** PMFs for AFP stacked onto graphene with protein sliding enabled (a) and disabled (b) at temperatures of 240, 260, 280 and 300 K. Gray dashed lines mark zero PMFs.

To capture the underlying contributions of protein sliding to free energy profiles during the formation of P-G interfaces, we investigated the potential of mean forces (PMFs) between the parallel AFP and graphene which allow free migration of protein in *x-y* plane. Similar to the stacking assembly of two graphene nanoplates,[40] the reaction coordinate (RC) of a stacking assembly was the separation between protein and graphene across the interface (*z*-direction). We found that several free energy barriers and traps must be overcome for the planar AFP to stack onto graphene along the RC (**Figure 6a**), especially at temperatures of 280 K, 260 K and 240 K. After AFP stacked onto graphene (RC < 0.5 nm), a clear free energy trap was observed. The depth of the free energy trap reflects the stability of the AFP adsorbing onto graphene, which decreased as the cooling of system and disappeared at temperatures lower than 260 K.

However, such a temperature-dependency of the depth decrease of the free energy traps was coincided with both the shrink of migration range of protein on graphene and the slow-down of water self-diffusion. To distinguish the influence of the transversal migration of protein to the free energy profiles from the thermal motion of water itself at different temperatures, other four PMFs were calculated using the identical umbrella potential along the same RC but with an additional restraint applied to the movement of AFP in the *x*- and *y*-directions (**Figure 6b**). We found that the depth of the free energy trap driving the adsorption of AFP onto graphene decreased dramatically as compared to the depth of the free energy trap for sliding enabled at 300 K. It demonstrated that the transversal migration of protein promoted the search for a suitable binding status of AFP on graphene. This indicates that water-stimulated protein sliding contributed mostly to protein binding on a water/graphene interface as compared with the thermal motion of water itself. In addition, when the sliding of AFP is disabled, the high energy barriers at an RC of approximately 0.5 nm present at all temperatures tested (300, 280, 260 and 240 K), confirming that the transversal motion of AFP could reduce the hindrance caused by the interlayer water, allowing protein to bind onto graphene.

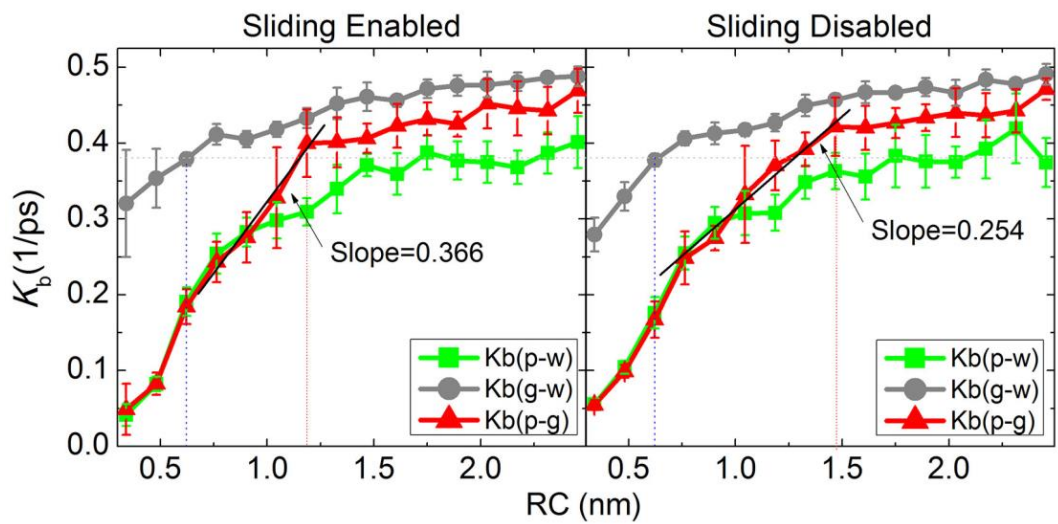

**Figure 7.** Changes in $K_b$(g-w), $K_b$(p-g) and $K_b$(p-w) with respect to RC when protein sliding was enabled or disabled.

To understand how the transversal migration of AFP reduces the hindrance caused by the interlayer water, the HB kinetics of water were further explored by monitoring the rate constant of HB breakage,[41] $K_b$, for the interlayer water ($K_b$(p-g)), the water on the surface of protein ($K_b$(p-w)), and the water on graphene ($K_b$(g-w))) (**Supplementary Figure S7**). Results showed that the values of $K_b$(p-g), $K_b$(p-w) and $K_b$(g-w) decreased when RC was reduced (with or without protein sliding) (**Figure 7**). When the water layers on protein and graphene overlapped with each other, the $K_b$(p-g) increased in a manner that was dependent on protein sliding. The slopes of $K_b$(p-g) for protein sliding enabled and disabled were approximately 0.366 and 0.254, respectively. $K_b$(p-g) was dependent on sliding at approximately 1.2 and 1.5 nm (vertical dotted red lines). The changes in $K_b$(p-g) over time suggest that protein sliding exerted its influence on HBs before the water layers on protein and graphene contacted with each other. However, $K_b$(g-w) and $K_b$(p-w) were not affected significantly by protein sliding until RC < 0.6 nm (marked with vertical dotted blue lines in **Figure 7**).

At RC = 0.6 nm, a monolayer of water was retained in the interlayer (knitted by intermolecular HBs, **Supplementary Figure S8**) within the P-G interface. Due to the strict P-G confinement when RC < 0.6 nm, the average values of $K_b$(p-g), $K_b$(p-w) and $K_b$(g-w) were further reduced either protein sliding was enabled or disabled. However, fluctuations (error bars) in $K_b$(g-w) and $K_b$(p-g) were higher when protein sliding was enabled than when protein sliding was disabled. This result implies that the last layer of water in the interlayer was drained and replaced with interfacial water on the graphene surface, confirming a faster breakage of intermolecular HBs between protein-coupled water and the non-coupled interlayer water induced by protein sliding relative to graphene.

## DISCUSSION

Based on integrated bio-nano interfaces constructed from the β-sheet model protein stacked on a graphene monolayer, we have explored the atomic-level interplay between a Thr abundant β-sheet protein, graphene, and water molecules within a P-G interface. The hindering role of the interlayer water on the stacking of the planar β-sheet protein with graphene was observed and coincided with changes in $E_{P-G}$ and $N_{IW}$. The residual water molecules within the interlayer impaired the $E_{P-G}$ even after the stacking of the protein onto graphene. By calculating the contribution of each protein residue to $E_{P-G}$, we found that only residues on protein surface (mostly were the Thr) dominated the assembled bio-nano interface. This finding allows us focused on the role of interlayer water on interaction between graphene and these contributive protein groups.

The interlayer water molecules filled in the interstices among the Thr arrays on protein surface, and could influence the dynamics behaviors of these contributive residues at molecular bonds level. The suppressed stretch vibrations of C-H molecular bonds on protein surface were caused by not only the P-G interaction but also the steric hindrance of interlayer water filled in the inter-group interstices on protein. Meanwhile, an obvious blue-shift of vibration mode around 3750 cm$^{-1}$ indicated a strong P-W intermolecular coupling inside the interface. Further, NCIs between protein and graphene identified via RDG analyses confirmed that the primary bio-nano attraction was the CH···π interaction between the methyl groups on Thr residues and the $sp^2$-carbon atoms on graphene. The RDG analyses also showed that two types of interlayer water molecules, *i.e.*, the protein-bound and unbound water molecules, could be distinguished by their differing protein-coupling characteristics. It means that blue-shift of VPS around 3775 cm$^{-1}$ can be attributed to the binding of interlayer water molecules on hydroxyl groups of Thr. From the transition of the intermolecular coupling between protein-bound and unbound water at temperatures of 300 K and 250 K, we found that the CH···π attraction between protein and graphene could be regulated by the HBs between water molecules in the interlayer.

On the other hand, the involvement of water molecules in the P-G stacking could active the transversal migration of protein upon graphene with a character of Levy flight. The interlayer water contributed to a larger walking range of AFP on graphene, but it could not change the character of Levy flight. In contrary to the interlayer water inside the P-G interface, the ripple of graphene and the slow-down of water diffusion could influence the Levy flight characteristic of protein upon graphene. Therefore, the environmental water and the confined water in the P-G interlayer both contributed to the unanchored adsorption of the β-sheet protein at the water/graphene interface. By calculating the free energy profiles of the stacking assembly of parallel protein and graphene at four tested temperatures (300 K, 280 K, 260 K, and 240 K), we found that transversal migration of protein promoted the searching for a suitable binding status of AFP on graphene because the depth of the free energy trap was related to the range of protein sliding. By disabling the migration of protein upon graphene in *x*- and *y*-directions manually, the influence of the transversal sliding of protein towards the free energy profiles could be wiped out. The contribution of the hindrance caused solely by the interlayer water to PMFs showed higher free energy barriers, especially when the protein closed to the graphene (RC~0.5 nm). These PMFs not only demonstrated the hindrance of interlayer water against the formation of bio-nano interface during the adsorption of a bulk β-sheet protein on graphene surface, but also revealed a physical basis of the spontaneous transversal migration of protein relative to graphene.

Recent calculations have demonstrated that the hindrance of interlayer water between two planar molecules such as graphene was caused by the confinement of two extended surfaces and the intermolecular coupling of water molecules themselves.[29] However, it is interesting that the sliding-enabled PMFs of the aggregate between parallel AFP and graphene were similar with those PMFs between two parallel graphene nanosheets without relative sliding.[40] Especially, relatively higher free energy barriers on correspondingly sliding-disabled PMFs revealed the contributions of P-W interaction on the stacking assembly of protein on graphene. Therefore, different with two smooth hydrophobic graphene surfaces, the P-W interaction along with the P-G confinement and the G-W interactions all can synergetic influence the behavior of interlayer water during the formation of an integrated bio-nano interface between protein and graphene. If the protein sliding associated with the motion of its hydration water, just like a scenario that a hydrated hydroxyl-rich β-sheet protein can be considered as a single entity competing with water molecules on an Au surface,[15] the relative sliding of the hydration water molecules on protein and graphene surfaces can lead to an unstable interlayer structure which subsequently eliminates its water molecules.

The investigation of HB dynamics for the water molecules within P-G interfaces demonstrated such a protein sliding-induced unstable intermolecular coupling of water. Within the P-G interface, the high HB breaking rate of the water layer on graphene showed an inherent restlessness of water molecules on a hydrophobic surface. It might be an origin of the observation that interlayer water can promote the transversal migration of protein on graphene. Oppositely, for the binding of water on protein surface groups (*e.g.* hydroxyl) and the P-W coupling which mediated by these protein-bound water molecules, HBs within the hydration layer were stable on the surface of protein. Although the water appeared to be confined to two hydration layers in contact with each other, the transversal motion of protein altered the HB breaking rate of the interlayer water. The splitting of the final layer of water between the protein and graphene corresponded to the highest penalty of free energy on the sliding-disabled PMFs, indicating that it was a crucial step in P-G aggregation. As expected, the fluctuation of the HB breaking rate of this water layer increased when protein sliding was enabled. These results demonstrated that when two hydration layers approached one another, the confinement of protein and graphene, the relative sliding between protein and graphene, the coupling of protein and water, and the interaction between graphene and water all modulated the intermolecular behaviors of the water molecules within the interlayer, determining the hindrance of interlayer water against the P-G assembly.

The regulatory mechanisms of water molecules in the assembly of a bulk β-sheet protein on graphene surfaces are more complex than that in the adsorption of small peptides on an extended solid surface. Although the interlayer water could hinder the assembly of both a bulk protein and a small peptide on the graphene-like surface, the interfacial-reoriented water layers preventing the contact between small peptides and the extended solid surface [12] would be destroyed by the transversal sliding of a bulk β-sheet protein upon graphene, but filled in the interstices among surface groups on the adsorbed-protein. Importantly, these interstitial water molecules still play roles in the specific interactions between an adsorbed-protein and graphene, determining the stability of the bio-nano interface. Meanwhile, we found that the hindering role of these interstitial water molecules in the assembly of proteins and hydrophobic graphene was different with that of peptides adsorption on hydrophilic surfaces such as Au and Pt.[15,42] The competing adsorption between hydrated peptides and water molecules on these hydrophilic materials,[15,42] has not been observed on a hydrophobic graphene surface. The physical basis of the regulatory role of these interstitial water molecules inside the hydrophobic P-G interface was originated from the intermolecular coupling of the confined interlayer water, which was mediated by the binding water molecules on protein surface. By cooling the system, we found that the intermolecular coupling of water on protein surface could control the adsorption of protein on graphene. Similarly, the interactions between water molecules and graphene would be enhanced effectively by tuning the charge density of the carbon atoms on graphene,[43] supplying a potential route in controllable assembly of proteins on graphene, particularly in the development of future single-molecule bio-nano devices.

In summary, we propose that the formation of a bio-nano interface by the adsorption of bulk β-sheet proteins on hydrophobic graphene surface in water environment will be regulated by the hindrance of interlayer water in both aspects of their interfacial interactions and their assembly process those causing traps and barriers on free energy profiles. Our findings should be lawful for the assembly of other bulk proteins owning a stable structure on graphene surface because most of the stable proteins in water have a strong coupling with water molecules on their surfaces. It is an important step towards a fundamental understanding of protein interactions with graphene in biological environments and in graphene-based biomedical devices, and can also enlighten the atomic-level exploration of bio-nano assembly for designing biocompatible materials by using proteins and other nanomaterials such as mesoporous carbon.

## METHODS

*Set-up of the Simulation System.* The initial structure of the model protein was obtained from the crystal structure of AFP in the Protein Data Bank (PDB entry 4DT5). To simulate the infinite graphene, a unit of graphene with a size of approximately $6.3 \times 6.3$ nm$^2$ (equal to the *x-y* box dimensions) was aligned along the *x-y* plane with the terminal carbon atoms sharing a chemical bond.[43-45] The graphene was positioned at the bottom of a $6.3 \times 6.3 \times 5.0$ nm$^3$ box. For flexible graphene, only carbon atoms at the cell edge were restrained to maintain position during simulations (**Supplementary Figure S9**). For rigid graphene, all graphene atoms were restrained to maintain smoothness of surface. The model protein was placed on top of the graphene sheet with an initial separation of approximately 1.5 nm. The simulation box was filled with water molecules to explore P-G assembly in water. Six chloride ions were added to the water box to neutralize the system. AFP was free to search for an appropriate assembly pathway in both water and in vacuo.

*Stability of the Stacking Conformation of AFP on Graphene.* To exclude the influence of stacking conformation stability of AFP on graphene, DSSP was utilized to analyze secondary structures of the model protein before data processing.[46] The evolution of the secondary structure of the protein over time was analyzed by evaluating MD trajectories of AFP adsorbing onto a graphene sheet in water and in vacuo. As a comparison, a 50 ns MD simulation of free AFP within a water box of $6.3 \times 6.3 \times 3.5$ nm$^3$ was performed to investigate the evolution of the secondary structure of the free protein over time. Compared with the stable secondary structure of the free AFP in water (**Supplementary Figure S10**), a small amount of AFP β-sheet transformed into coils in the early stage of stacking on graphene (blue dotted box, **Supplementary Figure S10**) and returned to β-sheet in the later assembly stage (red dotted box, **Supplementary Figure S10**) in water. This result correspond well with the water-protected β-sheet secondary structure of a bio-nano interface consisting of a β-sheet protein stacked onto graphene.[31] By contrast, this reversible secondary structure transition was not observed in vacuo (yellow dotted box, **Supplementary Figure S10**). However, the secondary structures of the adsorbed protein could be maintained at a certain level in vacuo, demonstrating the stable stacking conformation of AFP on graphene both in water and in vacuo.

*VPS Calculation.* The power spectrum or spectral density of vibration was obtained from the Fourier transform of the velocity autocorrelation function:[47]

$$\text{VPS}(v) = \lim_{t \to \infty} \frac{1}{2kT} \int_{-\tau}^{\tau} Vaf(t) e^{-2\pi v t} \, dt \qquad \text{(Eq. 1)}$$

The molecular vibration studies of AFP in free and binding statuses were carried out using the equilibrium structures of the free AFP and AFP-graphene complexes in water, respectively. The investigation of VPS of the contributing Thr groups without interlayer water was based on the same equilibrium system of AFP-graphene complexes, but the water molecules inside the interval of the P-G interface were manually deleted. Three independent MD simulations were carried out to obtain molecular vibrations of free-AFP in water and binding-AFP with and without the existence of interlayer water, respectively. In these simulations, the time step was set to 1 fs, the system coordinates and velocities were stored every 4 fs, and no constraints were applied to any molecular bonds. The VPS was computed using total of 125,000 structures (the last 500 ps in a 1 ns MD trajectory). The VPS of free AFP in a full frequency band was shown in **Supplementary Figure S3**. To identify the C-H stretching vibrations within the -$CH_3$ group of Thr residues, a benchmark MD simulation was performed for a poly-$(Thr)_3$ within a 3 nm cubic water box at 300 K. The obtained VPS is presented in **Supplementary Figure S4**.

*NCIs Analysis.* The NCIs involved in the P-G interface were presented by RDG(*r*):[34-36]

$$\text{RDG}(\boldsymbol{r}) = k \frac{|\nabla \rho(r)|}{\rho(r)^{4/3}} \qquad (\text{Eq. 2})$$

where ρ(r) is the electron density in real space and $k = 1/2(3\pi^2)^{1/3}$ is a constant.[35] The promolecular approximation was used to construct the electron density by superposing electron densities of free-state atoms.[36] The structures were sampled from a MD trajectory over 29 ns. Two 40 ns MD simulations (the first 11 ns were discarded) were performed at 250 and 300 K, respectively. In these simulations, the positions of both the protein and graphene were restrained, whereas the water molecules were free.

*PMF Calculation.* The umbrella sampling method was combined with the weighted histogram analysis method (WHAM) to calculate the PMFs.[48] The RC was defined as the distance between the graphene sheet and the $C_\alpha$ atoms of the protein along the normal direction of the protein-graphene interface. It needs to be noticed that the actual adsorption pathway of a protein on graphene surface is more complex than stacking along the normal direction of the interface. We focused our intention on the stacking assembly to understand the hindrance of interlayer water in this process. A 5 ns equilibrium simulation was performed before each sampling simulation to ensure the validity of the sampling (*e.g.,* the pressure balance of the interlayer water). Sixteen sampling simulations were carried out with the protein and graphene held constant at different protein-graphene distances from 2.5 nm (free status) to 0.3 nm (binding status). A biasing harmonic potential has been employed as the umbrella potential. The force constant of the umbrella potential was selected as 1,000 kJ (mol$^{-1}$nm$^2$) to engender a Gaussian distribution about the average value of the RC. In umbrella samplings of the sliding enabled PMFs, the movements of AFP in the *x*- and *y*-directions were not disturbed, as only a one-dimensional (1-D) umbrella potential was applied on the $C_\alpha$ atoms of the protein along the *z*-direction.[44] To avoid the sliding of the protein in the *x*- and *y*-directions, the heavy atoms in the protein were restrained by a 2D-harmonic potential with the same force constant as the umbrella potential. The umbrella positions were recorded at each time step (2 fs). MD trajectories with a total length of 3.4 μs were simulated to obtain all of the sampling data along the RC from approximately 0.3 to 2.5 nm.

*Study of the HB Dynamics.* Chemical dynamics were used to describe the kinetics of HB breakage based on the theory of Luzar and Chandler.[41] The HBs were defined based on the cutoffs for the acceptor-donor-hydrogen angle ($\alpha_{HB}$) and the acceptor-hydrogen distance ($r_{HB}$). The $\alpha_{HB}$ and $r_{HB}$ were chosen as 40° and 0.35 nm, respectively, to prevent missing HBs that could exist between water molecules. The HB dynamics of water was monitored along the RC of the PMFs.

*MD Parameters*. The AMBER99 force field was used to model proteins and ions.[49] The parameters for graphene carbon atoms are those of $sp^2$ carbon in benzene in the AMBER99 force field.[29] The TIP4P model was employed to describe water molecules.[50] The cut-offs of the vdW forces were implemented by a switching function starting at a distance of 1.1 nm and reaching zero at 1.2 nm. The particle mesh Ewald (PME) method was used to calculate the electrostatic interaction with a cut-off distance of 1.4 nm.[51] A stochastic velocity rescaling thermostat was used to control the temperature of the simulations.[52] Constraints were applied to all bonds to hydrogen with the LINCS algorithm,[53] which allowed for the use a 2 fs time step. A 3-dimensional harmonic potential with a force constant of 1,000 kJ (mol$^{-1}$ nm$^{-2}$) was used to restrain the coordinates of the carbon atoms on graphene. Unless otherwise specified, the energy of each simulation system was minimized for 200 steps; then, a 500 ps solvent relaxation process with the restraint of protein position was performed before the production simulations. Detailed descriptions of the simulations and calculation parameters are presented in **Supplementary Table S2**. All of the MD simulations were carried out using the GROMACS 4.5 software package.[54] All snapshots were rendered with VMD.[55]

# SUPPLEMENTARY INFORMATION

Detailed analysis of the R-G interaction; list of the production simulations and parameters; linear-fit plot of AFP adsorption onto graphene; plot of the monitored region for RDG analysis; plots of VPS of AFP and poly(Thr)$_3$; plots of the transversal shifts of AFP on graphene at different temperatures; plot of the hindering process of interlayer water on P-G stacking; plot of the final layer of water molecules inside the P-G interface; plot of the restrained atoms on the graphene edge; plots of the evolution of secondary structures of AFP; animation illustrating the MD trajectory.

# ACKNOWLEDGMENTS

The authors acknowledge the financial support from the National Natural Science Foundation of China (Nos. 21175134 and 21375125), the Creative Research Group Project of National Natural Science Foundation of China (No. 21321064).

# AUTHOR CONTRIBUTIONS

W.P.L. and R.A.W. conceived the concept. W.P.L. designed the study, performed the calculations, analyzed the data and wrote the manuscript, with contribution from R.A.W. G.J.X., H.Y.Z, X.L., S.J.L., and H.N. joined in the discussions with W.P.L. W.P.L. and R.A.W. revised the manuscript with inputs from X.D.S. and G.J.X.

# ADDITIONAL INFORMATION

Supplementary information accompanies this paper at http://www.nature.com/scientificreports

**Competing financial interests**: The authors declare no competing financial interests.